# Perfectly elastic collisions as the origin of the quantum states of superconductivity and magnetic order

Sh. Mushkolaj

## Abstract

One of the most interesting properties of solid materials is the ability to form different collective quantum states, such as superconductivity and magnetic order. This paper presents a model of perfectly elastic collisions (p.e.c.) as the universal origin of these collective quantum states. The superb agreement between calculated values and experimental data for critical temperatures, moreover, the explanation of the isotope effect in superconductivity and magnetic order confirms that this model successfully describes these two quantum states.

## Introduction

In many solid materials there are two or more phase transitions separated by their critical temperatures $T_c$. In each of these phases the behavior of the many-particle system is very different; for instance, in the quantum state of superconductivity, at temperatures $T<T_c$ the conduction electrons experience no electrical resistivity, above $T_c$ the electrical resistivity increases abruptly from zero to a non-zero value, which is called the normal conducting state. In the magnetically ordered states of ferromagnetism and antiferromagnetism, at temperatures below the Curie and Nèel temperatures, the electronic magnetic moments are aligned parallel and anti-parallel to each other: above $T_c$ they become disoriented. This state is called the paramagnetic state.

There are several theories that try to explain these collective quantum states. It is important to show how some of these theories calculate critical temperatures. In general, the equations for calculating critical temperatures are derived from the minimization of free energy. For instance, in the BCS theory [1] the minimization of free energy is done with respect to the distribution for electron pairs in ground states. The resulting equation for determining the critical temperature is:

$$T_c = \frac{1.14\hbar\omega}{k_B}\exp\left(-\frac{1}{N(0)V_{eff}}\right), \qquad (1)$$

where $\omega$, $N(0)$ and $V_{eff}$ represent the phonon frequency, the density of electronic states in the normal conducting state at Fermi level, and the effective electron-electron interaction, respectively. Because of exponential dependence, the effective electron-electron interaction $V_{eff}$ cannot be determined precisely enough to allow accurate computations of $T_c$. Another more sophisticated equation for $T_c$ is the so-called McMillan [2] equation derived from the Eliashberg theory [3], which is an extension of the BCS theory. Also, in the McMillan equation the parameter of the “renormalized” Coulomb repulsion is fairly arbitrary, and therefore the calculated values for $T_c$ may not be very accurate. For unconventional superconductors an appropriate theory able to derive an equation for accurate calculation of $T_c$ is still missing.

There are also several theories that attempt to explain the collective quantum state of the magnetic order. It is necessary to provide a short overview of the most accepted and successful models and theories, which attempt to describe this phenomena. There is a classification based on the type of magnetic moment, that can be either localized or itinerant [4]. The magnetic order of the localized moments appears in the insulators and is described quite well by the Heisenberg model [5]. The ordered state of the itinerant moments, called band magnetism is characteristic of the metallic systems; this phenomena is described by the Hubbard model [6]. There are also some systems, such as 4f-systems, where a combination of these two models is needed because the magnetic order and electrical conductivity are caused by two different electron groups.

Theoretical computations of the critical temperatures in many-particle systems is no easy task because one often needs to make many approximations and assumptions to get values that are in agreement with experimental results. It is worth mentioning the Weiss and Stoner models, which are the mean-field approximations of the Heisenberg and Hubbard models, respectively. From the Weiss model the critical (*i.e.*, Curie and Nèel) temperatures are expressed by:

$$T_c = JC, \tag{2}$$

where $J$ and $C$ represent the exchange parameter and Curie constant, respectively. From the Stoner model the following equation for $T_c$ results:

$$T_c = \frac{W}{4 k_B arctanh(W/U)}, \tag{3}$$

where $W$ represents the band width and $U$ the Coulomb repulsion. Equation (3) usually overestimates the $T_c$ and therefore it seems to be useless [4]. Another method to extract $T_c$, which has become quite popular in recent years, is the random phase approximation.

As one can see from Eq. (2) and (3) there is no direct dependence of the $T_c$ from the atomic-mass. The missing of the atomic-mass dependence eliminate the possibility to explain the isotope effect in magnetic order. In this paper the direct atomic-mass dependence of the $T_c$ in magnetically ordered systems is introduced.

Although superconducting and magnetically ordered states are are different, they are similar in two aspects: they are not destroyed by heat energy at temperatures below $T_c$, and they are quantum states. Based on these similarities two different methods are used to derive the equations for critical temperatures. The first method is based on the atom-atom and electron-atom perfectly elastic collisions, in superconducting and magnetically ordered states, respectively. The second method is based on the solution of the time dependent Schrödinger equation, where atom and electron eigenstates are plane waves.

In the following sections it will be shown that the model of the p.e.c. has a universal character, where superconductivity and magnetic order are unified into one collective quantum state, where the uncertainty principle and the conservation of the kinetic energy are obeyed.

In the first section that is separated into five subsections, the quantum state of superconductivity is described. In the first and second subsections two methods are used to derive equations for $T_c$ , namely, the method of the p.e.c. and the method based on the solution of the time dependent Schrödinger equation, respectively. In the third subsection the isotope effect in superconductivity is treated. The Fourth and the fifth subsections deal with the Cooper pair formation and with the incorporation of the conduction charges into the perfectly elastic atom-atom collisions through the London penetration depth, respectively.

In the second section that is separated into three subsections the quantum state of the magnetic order is treated. In the first and second subsections two methods are used to derive equations for $T_c$ , namely, the method of the p.e.c. and the method based on the solution of the time dependent Schrödinger equation, respectively. The third subsection deals with the isotope effect in magnetic order.

In the third section the coexistence between quantum states of superconductivity and magnetic order is shortly discussed.

## 1. Superconductivity

There are two types of superconductors: conventional superconductors with lower critical temperatures and unconventional or "high-temperature" superconductors (H$T_c$ S) with higher $T_c$ . The properties of the conventional superconductors are described by the BCS theory [1]. The basic assumption of this theory is the pairing of conduction electrons into so-called Cooper pairs ($k$, -$k$), where the attractive force between two electrons results from the electron-phonon interaction. However, the BCS theory fails to explain the properties of the quantum state of H$T_c$ S, thereby puzzling theoreticians and experimentalists for a long time.

As it will be demonstrated below, the formation of the state of pairing wave vectors ($k$, -$k$) (*i.e.* Cooper pairs) follows directly from the perfectly elastic collisions. In the following pages of this section is shown also that classification of superconductors as conventional or unconventional is not necessary because in both cases the origin of superconductivity is actually the same. Namely in both types of superconductors the p.e.c. between atoms or between groups of atoms are the fundamental source for creating this quantum state.

In the subsection 1.5. is shown that in addition to the p.e.c. between atoms there exist also p.e.c. between conduction charges (*i.e.* electrons or holes) and atoms. Based on the conservation of the kinetic energy, the connection between the wave vectors of the conduction charges ($k_{c.c.}$ ) and the phonon wave vectors ($k$) is given by the formula $k_{c.c.}=(m_e/M)^{1/2}\, k$.

### 1.1. The method of the p.e.c.

In the collective quantum state of superconductivity there exist atom-atom collisions which are perfectly elastic, *i.e.* from these collisions no kinetic energy is transformed into heat energy. By following this idea the equations for $T_c$ for conventional and unconventional superconductors are derived.

During a round trip of two p.e.c. each center of atomic mass (nucleus) in a crystal can travel a distance of d = 2($a$ -$\varepsilon$), where $a$ is the equilibrium atom-atom bond length and $\varepsilon$ is the nucleus-nucleus distance during a collision. As demonstrated in Figure 1a, for atoms with a zero displacement of the positive (nucleus) and negative (electron) charge centers, the nucleus-nucleus distance during a collision is $\varepsilon$ =2R, where R is the atomic radius. However, during atom-atom collisions, the positive and negative charge centers may experience a displacement as shown in Figures 1b and 1c for ε equal to R and R/2, respectively. As it will be shown later the agreement between experimental and calculated data is reached for values of ε =2R, R and R/2. In more details the ε parameter will be explained in the Appendix.

Fig. 1 shows that in the one-dimensional atomic chain of N atoms with atomic mass *M* at the time $t= t_0$, the first atom starts to move with a velocity V. Due to momentum conservation, after the first collision at $t=t_1$ the velocity of the first atom jumps to zero but the second atom starts to move with the same velocity until $t=t_2$ when it collides with the third atom, and after the collision with the third atom the velocity of the second atom jumps to zero, and so forth. After 2N atom-atom collisions at time $t=t_{2N}$, all atoms return again to their initial positions. Because collisions are perfectly elastic the absolute value of the velocity and the time difference between collisions $\Delta t/2=t_i-t_{i-1}$ for *i*= 1, 2,…2N, remain constant. The time $\Delta t/2$ needed for each nucleus to travel the distance $(a - \varepsilon)$ is given as:

$$\frac{\Delta t}{2}=\frac{(a-\varepsilon)}{V}. \tag{4}$$

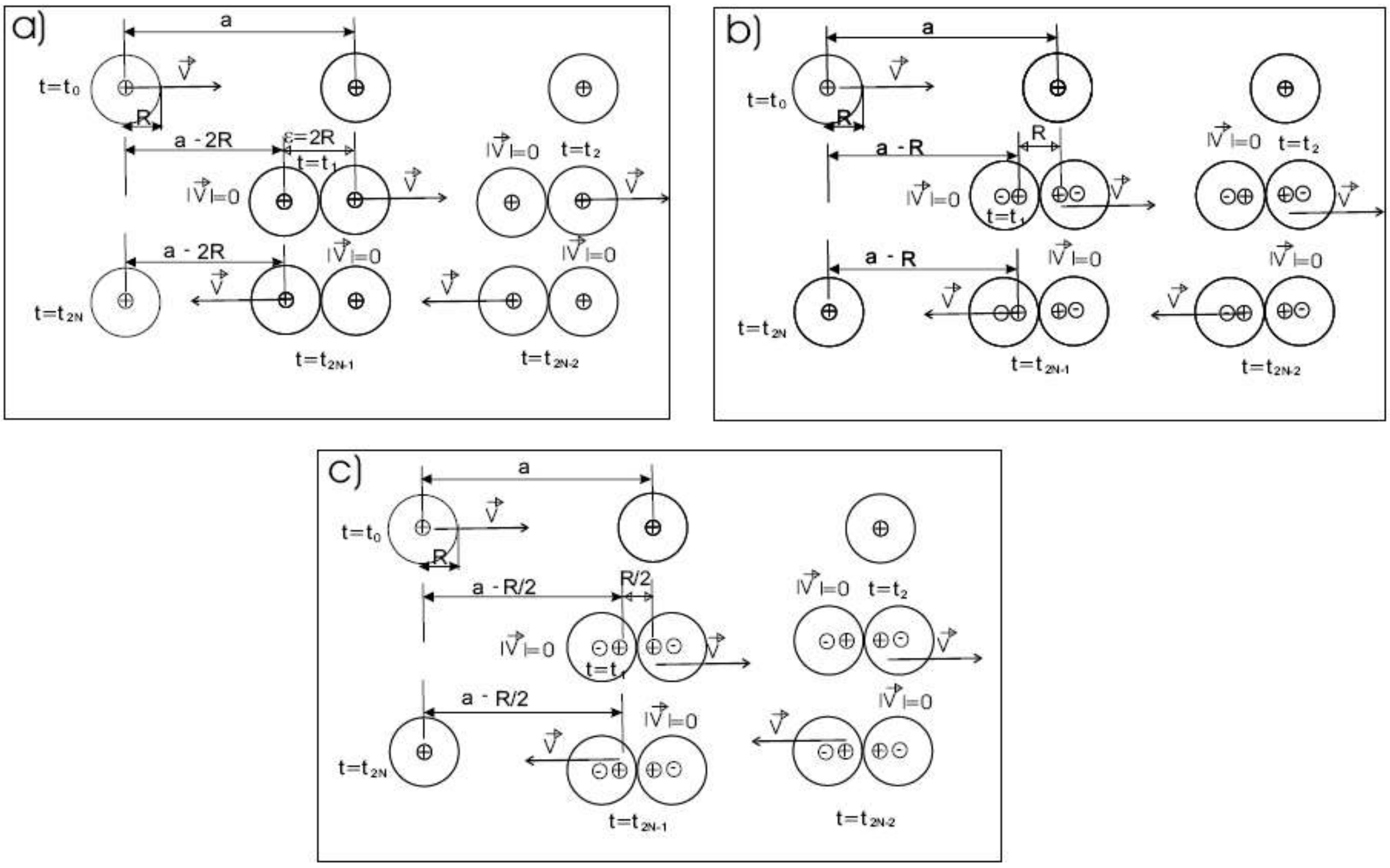

*Figure 1: Figure a represent the case when atoms behave like hard spheres and therefore during collisions the distance between two nuclei is ε=2R. In the Figures b, and c atoms behave like "soft" spheres and therefore during collisions the distances between two nuclei are ε=R and ε=R/2, respectively. Because atom-atom collisions are perfectly elastic the absolute value of the velocity V and the time difference between collisions* $\Delta t/2=t_i-t_{i-1}$ *for i= 1, 2,…2N, in all three cases (i.e. ε=2R, ε=R and ε=R/2 ) remain constant.*

In the following pages $\Delta t$ represent the time difference between two perfectly elastic collisions.

During the time $\Delta t$ the atomic translational momentum $MV$ is equal to the crystal momentum $hk$ ; *i.e.*:

$$MV = hk, \tag{5}$$

where $k = 2\pi/d$. Inserting the atomic velocity $V$ from Eq. (5) into Eq. (4), the time difference between two p.e.c. ($\Delta t$ ) gives:

$$\Delta t=2M\frac{(a-\varepsilon)^2}{\pi h}. \tag{6}$$

Every quantum state is characterized by the time-energy uncertainty relation with the formula $\Delta t\, \Delta E = h$. During the time $\Delta t$ the kinetic energy for each atom jumps between only two values, namely $E_k = 0$ and $E'_k = k_B T_c$. For $\Delta E = E'_k - E_k = k_B T_c$ one gets

$$\Delta t\, k_B T_c = h. \tag{7}$$

By inserting $\Delta t$ from Eq. (6) into Eq. (7) the equation for critical temperature $T_c$ adopts the form:

$$T_c=\frac{\pi h^2}{2k_B}\frac{1}{M(a-\varepsilon)^2}. \tag{8}$$

According to Eq.(8), the critical temperature depends only on the atomic mass and on the square of the difference between the equilibrium atom-atom bond length $a$ and the nucleus-nucleus distance ε during a collision. The calculated values for $T_c$ of different superconducting materials compared with the experimental data are listed in Table 1.

The agreement between calculated and experimental data is spectacular. Most interesting cases are those of binary, ternary and quaternary compounds where one can deduce which type of atoms cause the appearance of superconductivity. However, in some compounds such as, $CaSi_2$, $KOs_2O_6$, 2H-$NbSe_2$, $CeCoIn_5$, $YPd_2B_2C$, $ThPd_2B_2C$, $YNi_2B_2C$, $LuNi_2B_2C$ and $La_3Pd_4Si_4$ there may exist two or more types of atoms which contribute to superconductivity.

At this point it is also worth to emphasize the fact that atoms which behave like “soft” spheres (*i.e.* ε =R and ε =R/2 ), are located at the sites with non-cubic symmetry, where large electric field gradients (EFG) are expected. Therefore, detailed NMR and NQR experiments could give more insights about the ε parameter.

| Material, Struc. | ε | a(Å) | R(Å) [29] | $T_c$ (K) calc. | $T_c$ (K) exp. |
|---|---|---|---|---|---|
| $V_3Si$, A15 | $2R_{Si}$ | a=4.7235 | 1.1 | 16.6 | 16.5 [7] |
| Pb, FCC | 2R | a=4.95 | 1.80 | 7.9 | 7.2 [8] |

Continuation of Table 1.

| | | | | | |
|---|---|---|---|---|---|
| $LiTi_2O_4$, CUB | $2R_{Li}$ | a=8.4 | 1.45 | 14.2 | 13 [9] |
| La, HEX | 2R | c=6.07 | 1.95 | 4.56 | 4.88 [8] |
| $Ba_{0.6}K_{0.4}BiO_3$, CUB | $R_O$ | 0.707*4.293 | 0.6 | 31.4 | 30 [10] |
| $PuCoGa_5$, TET | $R_{Ga}$ | 0.707*a=2.774 | 1.3 | 19.7 | 18.5 [11] |
| $Li_{0.16}ZrNCl$, layered | $R_{Zr}$ | $d_{Zr\text{-}Zr}$=3.1 | 1.55 | 13.6 | 15 [12] |
| $BaPb_{0.75}Bi_{0.25}O_3$, ORTH | $R_O$ | c/2=4.25 | 0.6 | 14 | 13 [13] |
| $Nb_3Ge$, A15 | $R_{Nb}$ | $d_{Nb\text{-}Nb}$=2.589 | 1.45 | 24.7 | 23.2 |
| $CaSi_2$, $AlB_2$-type above 16 GPa | $R_{Si}$ | a=3.7077 | 1.1 | 15.5 | ≈14 [14] |
| $CaSi_2$, $AlB_2$-type above 16 GPa | $R_{Ca}$ | c=4.0277 | 1.8 | 15 | ≈14 [14] |
| $ScNi_2B_2C$, TET | $R_{Ni}$ | 3.37 | 1.35 | 12.5 | 13.5 [17] |
| $LiTi_2O_4$, CUB | $R_O$ | a/2=4.2 | 0.6 | 14.3 | 13 [9] |
| $CaC_6$ | $R_{Ca}$ | a=4.333 | 1.8 | 11.6 | 11.5 [15] |
| $Li_3Ca_2C_6$ | $R_{Li}$ | a=7.45 | 1.45 | 11.9 | 11.15[16] |
| $ErNi_2B_2C$, TET | $R_{Ni}$ | 3.495 | 1.35 | 11 | 11 [17] |
| $TmNi_2B_2C$, TET | $R_{Ni}$ | 3.494 | 1.35 | 11 | 11 [17] |
| $NbB_2$, HEX | $R_{Nb}$ | c =3.267 | 1.45 | 9.7 | 9.75 [18] |
| $KOs_2O_6$, CUB | $R_K$ | a/2=5.05 | 2.2 | 9.4 | 9.6 [19] |
| $KOs_2O_6$, CUB | $R_O$ | a/2=5.05 | 0.6 | 9.4 | 9.6 [19] |
| $KOs_2O_6$, CUB | $R_O$ | $d_{O1\text{-}O1}$=2.819, n=4 [20] | 0.6 | 9.5 | 9.6 [19] |
| Nb, BCC | R | a=3.30 | 1.45 | 9.4 | 9.25 [8] |
| $MgCNi_3$, CUB | $R_{Ni}$ | 3.82 | 1.35 | 8.3 | 8.5 [21] |
| 2H-$NbSe_2$, HEX | $R_{Nb}$ | a=3.443 | 1.45 | 8 | 7.2 [22] |
| 2H-$NbSe_2$, HEX | $R_{Se}$ | a=3.443 | 1.15 | 7.2 | 7.2 [22] |
| $YB_6$ , CUB | $R_Y$ | a=4.1 | 1.8 | 6.3 | 6.5-7.2 [23] |
| $RbOs_2O_6$, CUB | $(R_O+R_{Rb})/2$ | a/2=5.057 | 0.6; 2.35 | 6.3 | 6.3 [24] |
| La, FCC | R | 0.707*a=3.754 | 1.95 | 6.6 | 6.06 [8] |
| $DyNi_2B_2C$, TET | $R_{Dy}$ | 3.534 | 1.75 | 5.8 | 6 [17] |
| SrAlSi, HEX | $R_{Sr}$ | c=4.754 | 2 | 4.5 | 4.9 [25] |
| Ta, BCC | R | a=3.31 | 1.45 | 4.77 | 4.48 [8] |
| Hg, RHL | R | $a=3\sqrt{2(1-\cos(70.52^{\circ}))}$ | 1.5 | 3.85 | 3.95 [8] |
| In, TET | R | a=4.59 | 1.55 | 2.8 | 3.4 [8] |
| $CsOs_2O_6$, CUB | $R_{Cs}$ | a/2=5.0745 | 2.6 | 3.6 | 3.3 [26] |

Continuation of Table 1.

| | | | | | |
|---|---|---|---|---|---|
| $CeCoIn_5$, TET | $R_{In}$ | a=4.613 | 1.55 | 2.7 | 2.3 [27] |
| $CeCoIn_5$, TET | $R_{Ce}$ | a=4.613 | 1.85 | 2.7 | 2.3 [27] |
| $CeCoIn_5$, TET | $R_{Co}$ | a=4.613, n=2 | 1.35 | 2.3 | 2.3 [27] |
| $OsB_2$, ORTH | $R_{Os}$ | c=4.0771 | 1.3 | 2.03 | 2.1 [28] |
| Re, HEX | R | c=4.45 | 1.35 | 1.66 | 1.69 [8] |
| Th, FCC | R | a=5.08 | 1.80 | 1.2 | 1.37 [8] |
| Ga, TET | R | a=7.6633 | 1.3 | 1.06 | 1.091 [8] |
| Gd, HEX | R | a=5.7826 | 1.8 | 1.19 | 1.083 [29] |
| UBe13, CUB | $R_U$ | $d_{U-U}$=5.13 | 1.75 | 1 | ≈0.9 [30] |
| U, ORTH | R | c=5.87 | 1.75 | 0.7 | 0.7 [31] |
| $Li_2Pt_3B$, CUB | $R_{Pt} + R_B/2$ | a=6.7552 | 1.35; 0.85 | 2.6 | 2.43 [32] |
| $MgB_2$, HEX | $R_B/2$ | a=3.082 | 0.85 | 39.1 | 39.4 [33] |
| $YPd_2B_2C$, TET | $R_B/2$ | 3.71 | 0.85 | 25.6 | 23.4 [34] |
| $YPd_2B_2C$, TET | $R_C/2$ | 3.71 | 0.7 | 22 | 23.4 [34] |
| $ThPd_2B_2C$, TET | $R_B/2$ | 3.839 | 0.85 | 23.7 | 20-21 [35] |
| $ThPd_2B_2C$, TET | $R_C/2$ | 3.839 | 0.7 | 20.4 | 20-21 [35] |
| $Na_{0.29}HfNCl$, layered | $R_N/2$ | 3.5892 | 0.65 | 20 | 20 [12] |
| $LuNi_2B_2C$, TET | $R_B/2$ | 3.46, n=2 | 0.85 | 15 | 16.6 [17] |
| $LuNi_2B_2C$, TET | $R_C/2$ | 3.46, n=2 | 0.7 | 12.8 | 16.6 [17] |
| $LuNi_2B_2C$, TET | $R_{Ni}/2$ | 0.707*a=2.446 | 1.35 | 16.3 | 16.6 [17] |
| $YNi_2B_2C$, TET | $R_B/2$ | 3.526, n=2 | 0.85 | 14.3 | 15.4 [17] |
| $YNi_2B_2C$, TET | $R_C/2$ | 3.526, n=2 | 0.7 | 12.3 | 15.4 [17] |
| $YNi_2B_2C$, TET | $R_{Ni}/2$ | 0.707*a=2.493 | 1.35 | 15.4 | 15.4 [17] |
| $PuRhGa_5$, TET | $R_{Ga}/2$ | 0.707*a=2.99 | 1.3 | 7.8 | 8.5 [36] |
| $HoNi_2B_2C$, TET | $R_{Ni}/2$ | 0.707*a=2.487, n=2 | 1.35 | 7.7 | 8 [17] |
| CaAlSi, HEX | $R_{Si}/2$ | a=4.189 | 1.1 | 8 | ≈7.9 [25] |
| CaAlSi, HEX | $R_{Al}/2$ | a=4.189 | 1.25 | 8.7 | ≈7.9 [25] |
| Tc, HEX | R/2 | a=2.735 | 1.35 | 7.2 | 7.7 [8] |
| $Li_2Pd_3B$, CUB | $R_B/2$ | a=6.753 | 0.85 | 6.9 | ≈7.9 [37] |
| $YbC_6$ | $(R_{Yb}+R_C)/2$ | a=4.32 | 1.75; 0.7 | 6.7 | 6.5 [38] |
| $ZrB_{12}$ | $R_B/2$ | a=7.4 | 0.85 | 5.6 | 6 [39] |
| $YB_{12}$ | $R_B/2$ | a=7.5 | 0.85 | 5.5 | 4.7 [40] |
| Sn, TET | R/2 | c=3.182 | 1.45 | 4.1 | 3.72 [8] |

Continuation of Table 1.

| | | | | | |
|---|---|---|---|---|---|
| $Na_{0.35}CoO_2 1.3H_2O$ | $R_{Co}/2$ | 1.414*a=3.992 | 1.35 | 4.6 | 4.54 [41] |
| $CdCNi_3$ | $R_{Cd}/2$ | a=3.844 | 1.55 | 2.8 | 2.5-3.2 [42] |
| $La_3Pd_4Ge_4$, ORTH | $R_{Ge}/2$ | b=4.385 | 1.25 | 2.9 | 2.75 [43] |
| Tl, HEX | R/2 | a=3.46 | 1.90 | 2.1 | 2.39 [8] |
| $La_3Pd_4Si_4$, ORTH | $R_{La}/2$ | a=4.22 | 1.95 | 2.04 | 2.15 [44] |
| $La_3Pd_4Si_4$, ORTH | $R_{Pd}/2$ | a=4.22 | 1.4 | 2.25 | 2.15 [44] |
| $PrOs_4Sb_{12}$ | $(R_{Os}+R_{Sb})/2$ | a/2=4.6525 | 1.3; 1.45 | 1.85 | 1.85 [45] |
| $RuB_2$, ORTH | $R_{Ru}/2$ | a=4.6457 | 1.3 | 1.8 | 1.6 [28] |
| Sr2RuO4 | $R_{Sr}/2$ | 1.414*a=5.47 | 2 | 1.7 | 1.5 [46] |
| Pa, TET | R/2 | a=3.92 | 1.80 | 1.4 | 1.4 [8] |
| Y, HEX | R/2 | c ≈5.734 | 1.80 | 1.4 | 1.3 [29] u. pressure |
| UGe2, ORTH | $R_U/2$ | c= 4.116 | 1.75 | 1.1 | 1 [47] |
| UPt3, $MgCd_3$-type | $R_U/2$ | a= 5.764 | 1.75 | 0.51 | 0.53 [48] |
| URhGe, ORTH | $R_U/2$ | c= 7.51 | 1.75 | 0.28 | 0.25 [49] |
| Cd, HEX | $R_{Cd}/2$ | c= 5.618, n=2 | 1.55 | 0.56 | 0.56 [8] |
| ZrZn2, CUB | $R_{Zr}/2$ | 1.414*a=10.455 | 1.55 | 0.3 | 0.29 [50] |

*Table 1. Calculated $T_c$-s for different superconductors are compared with experimental values.*

Under pressure the $T_c$ for fcc Lanthanum rises very rapidly from 6 K at ambient pressure and saturates at nearly 13 K around 200 kbars [51]. The volume also decreases rapidly with increasing pressure [52].

In Table 2 the calculated values for $T_c$ at different pressure values are listed .

| P (kbar) | $V/V_0$ [52] | 0.707*a (Å) | $T_c$ (K) calcul. | $T_c$ (K) calcul. [51] | $T_c$ (K) exp. |
|---|---|---|---|---|---|
| 0 | 1 | 3.754 | 6.6 | 8.3 | 6.06 |
| 50 | 0.85 | 3.556 | 8.33 | 11.6 | 10.0 |
| 120 | 0.733 | 3.385 | 10.44 | 14.5 | 11.6 |
| 207 | ≈0.65 | 3.252 | 12.7 | -- | ≈12.8 |

*Table 2. Calculated and experimental data for $T_c$-s of fcc Lanthanum at different pressure values. The data for $V/V_0$ at 50 and 120 has been taken from ref. [52]. The value for $V/V_0$ at 207 kbars has been deduced by using the Murnaghan equation.*

The value for $V/V_0$ at 207 kbars has been calculated by inserting the same values for the initial isothermal bulk modulus of 248 kbar and pressure derivative of 2.8 into the Murnaghan equation presented in reference [52]. The calculated data from Table 2 for fcc

Lanthanum confirms that the model of p.e.c. is also able to describe the increasing of the critical temperature with increasing pressure.

On the first view Eq. (8) fails to calculate the $T_c$-s for some superconductors such as V, Zr and Al. For example, by calculating the $T_c$-s from Eq. (8) with $\varepsilon = R$ one gets values of 20.75 K, 2.53 K and 14.1 K for V, Zr, and Al respectively, which are larger than the experimental values of 5.4 K, 0.61 K and 1.17 K. To achieve the agreement with experimental values, it is necessary to consider the fact that beside the collisions between single atoms, collisions between groups of *n* simultaneously vibrating atoms may exist. As it is shown above in Table 1 for some compounds the "two-atoms"-"two-atoms" collisions (*i.e., n*=2) are introduced. The momentum of a group of *n* simultaneously vibrating atoms is given as *nMV.* In this case Eq. (5) adopts the form:

$$nMV = hk\,, \tag{9}$$

and the general equation for the critical temperature is given by:

$$T_c = \frac{\pi h^2}{2k_B}\frac{1}{nM(a-\varepsilon)^2}. \tag{10}$$

For *n*=1 Eq. (10) transforms into Eq. (8). For n=4, 4 and 10 the previously calculated values from Eq. (8) for V, Zr, and Al change to 5.2 K, 0.63 K and 1.1 K, respectively, which are in good agreement with experimental values.

In unconventional superconductors the p.e.c. that induce the collective quantum state of superconductivity are not between two atoms of the same atomic mass as in the case of conventional superconductors where atoms of the same sort collide with each other.

First, the unconventional superconductor $Yba_2Cu_4O_8$, where p.e.c. are between Cu and O atoms, will be analyzed. The radii for Cu and O atoms are $R_{Cu}=1.35$ *Å* and $R_O=0.6$ *Å* [29], and their masses are $M_{Cu}=106.2937\ 10^{-27}$ *kg* and $M_O=26.7599\ 10^{-27}$ *kg* respectively. During the perfectly elastic Cu-O collisions kinetic energy is conserved; *i.e.*,

$$\frac{P_O^2}{2M_O} = \frac{P_{Cu}^2}{2M_{Cu}}. \tag{11}$$

For $P_O=M_OV$ and $P_{Cu}=hk$, or vice versa, $P_{Cu}=M_{Cu}V$ and $P_O=hk$, Eq. (11) transforms into:

$$\sqrt{M_{Cu}M_O}\,V = hk\,, \tag{12}$$

where $k=2\pi/2(\alpha-\varepsilon)$ and $\alpha$ is the average bond length between Cu and O*.* The time $\Delta t/2$ needed for each atom to travel the inter-atomic distance of $(\alpha-\varepsilon)$ is given as:

$$\frac{\Delta t}{2} = \frac{(\alpha-\varepsilon)}{V}. \tag{13}$$

Inserting velocity $V$ from Eq. (12) into Eq. (13), the time difference between two p.e.c. $\Delta t$ get the form:

$$\Delta t = 2\sqrt{M_{Cu} M_O}\frac{(\alpha-\varepsilon)^2}{\pi h}. \qquad (14)$$

Using Eq. (7) for time-energy uncertainty $T_c$ becomes:

$$T_c = \frac{\pi h^2}{2k_B}\frac{1}{\sqrt{M_{Cu} M_O}(\alpha-\varepsilon)^2}. \qquad (15)$$

Thus for $M_{Cu}$=$M_O$, Eq. (15) is transformed into Eq. (8).

By applying pressure up to 12 GPa the $T_c$ of $YBa_2Cu_4O_8$ increases from 80 K to the saturated value of 108 K, where saturation begins above the pressure of 5 GPa [53].

There are two reasons why $T_c$ for $YBa_2Cu_4O_8$ increases with increasing pressure: first because by increasing the pressure the average minimal Cu-O bond length $\alpha$ decreases, and second because p.e.c. change direction with increasing pressure.

Table 3 provides some calculated values for the $T_c$ of $YBa_2Cu_4O_8$ for different pressure values and for $\varepsilon = (R_O+R_{Cu})/2$.

| Pressure | $a_{Cu\text{-}O}$(Å) | $T_c$ (K) calcul. | $T_c$ (K) exp.(ref[53]) |
|---|---|---|---|
| ambient-press. | c-direction<br>$(a_{Cu(2)\text{-}O(1)} + a_{Cu(1)\text{-}O(1)})/2$ =2.0645 (ref[53]) | 79 | 80 |
| 4.65 GPa | b-direction<br>$a_{Cu(2)\text{-}O(3)}$ = 1.947 (ref[53]) | 99 | ≈99 |
| 12 GPa>Pressure>5 GPa | a-direction<br>$a_{Cu(2)\text{-}O(2)}$ = 1.9 (ref[53]) | 109 | ≈108 |

*Table 3. Calculated $T_c$-s for the unconventional superconductor $YBa_2Cu_4O_8$ are compared with experimental data at different pressure values.*

It is evident that the agreement between calculated and experimental values listed in Table 3 is excellent. Now Eq. (15) will be tested to determine if it is also able to calculate the critical temperatures for other unconventional superconductors.

$La_2CuO_{4+\delta}$ compounds are orthorhombic and show an enhanced orthorhombic distortion with longer oxidation times. For short oxidation times this unconventional superconductor exhibits two superconducting transitions at about 32 K and 45 K [54]. The equilibrium Cu-O bond length $\alpha$ depends on the direction; *e.g.*, in *c* direction it is approximately 2.428 *Å* [55], and in *a* and *b* directions $\alpha$ is approximately $a/2 \approx 2.669$ *Å* and $b/2 \approx 2.701$ *Å*, respectively. Using the same value as for $\varepsilon = (R_O+R_{Cu})/2$ in $YBa_2Cu_4O_8$ one obtains critical temperatures of 44.4 K, 32.6 K, and 31.4 K in *c*, *a* and *b* directions, respectively, which are in excellent agreement with the experimental values of 32 K and 45 K. $YBa_2Cu_3O_7$ is a superconductor with a critical temperature of about 93 K. In the unit cell of this compound there are planar Cu-O distances of 1.930 *Å* and 1.964 *Å* [56]. Inserting the planar distance of 1.964 *Å* into Eq. (15) gives the value 95.7 K.

It is also interesting to point out that in some cases such as 2H-$NbSe_2$ both equations (namely 8 and 15) yield the same value for $T_c$. 2H-$NbSe_2$ has a hexagonal layered structure with lattice parameters of a= 3.443 *Å* and c= 12.547 *Å* [22]. Inserting the planar Nb-Se distance of 3.443 *Å* and $\varepsilon =(R_{Nb}+R_{Se})/2$= 1.3 *Å* into Eq. (15) the $T_c$ value becomes 7.5 K, which is in good agreement with the experimental value of 7.2 K.

It is very interesting to analyze the case of $Ba_{(1-x)}K_xBiO_3$ where two oxidations states of Bi namely, $Bi^{3+}$ and $Bi^{5+}$ [57] coexist. Because these two ions of Bi have different Radii, there are two unequal $Bi^{3+}O_6$ and $Bi^{5+}O_6$ octahedra, and two different Bi-O distances. At the temperature of 13 K for x=0.2 there are two Bi-O distances of 2.24 *Å* and 2.09 *Å* [57]. These two Bi-O distances are used to calculate the *Tc*-s for $Ba_{0.6}K_{0.4}BiO_3$. *Tc*-s of the order of 15 K, 20 K [58] and 29.8 K [59] have been measured. Except Barium all three other atoms contribute to the superconductivity. For K-K collisions, a= 2*2.24 *Å,* a= 2*2.09 *Å* and $\varepsilon =R_K$ one get values of 14.6 K and 19.5 K, respectively. For O-O collisions, a= 1.414*2.09 *Å,* a= 1.414*2.24 *Å,* a=2*2.09 *Å,* and $\varepsilon =R_O$ one get values of 33 K, 28.3 K and 14.5 K, respectively. For $Bi^{3+}$-O and $Bi^{5+}$-O collisions, a=2.24 *Å*, a=2.09 *Å*, $\varepsilon = (R_{Bi3+}+R_O)/2$ and $\varepsilon =(R_{Bi5+}+R_O)/2$ one get *Tc* values of 28.1 K and 28.7 K, respectively.

To date, the highest *Tc* value at ambient pressure has been measured in superconducting cuprate of $(Hg_{0.8}Tl_{0.2})Ba_2Ca_2Cu_3O_{8.33}$ which is 138 K, and it seems it is hopeless to find cuprate compounds with higher *Tc*. Why? Because in cuprates there are no shorter Cu-O distances than 1.8 *Å* [60] at ambient pressure. If the collisions between Cu and O atoms with shortest Cu-O distance of 1.8 *Å* are perfectly elastic and contribute to the superconductivity than from the Eq. (15) one get a maximal *Tc* of 137.5 K. Under pressure this bond length may be reduced and higher *Tc-s* can be achieved. However, it is very possibly that other materials with light atoms and short bond lengths can provide superconductivity with higher *Tc* -s than 138 K.

### 1.2.The method based on the solution of the time dependent Schrödinger equation

As it has been shown above the quantum state of superconductivity is caused due to the elastic collisions between atoms with masses $M_1$ and $M_2$, and respective kinetic energies of $E_{k1}=P_1^2/2M_1$ and $E_{k2}=P_2^2/2M_2$. During the perfectly elastic collisions, the kinetic energies are conserved, i.e. $E_{k1}=E_{k2}$. In this case the kinetic term of the Hamiltonian operator for a free "particle" with the mass of $\sqrt{M_1 M_2}$ may be expressed as:

$$\hat{H}=\frac{-\hbar^2}{2\sqrt{M_1 M_2}}\frac{\partial^2}{\partial x^2}. \tag{16}$$

The time-dependent Schrödinger wave equation is:

$$\frac{-\hbar^2}{2\sqrt{M_1 M_2}}\frac{\partial^2}{\partial x^2}\Psi(x,t)=i\hbar\frac{\partial}{\partial t}\Psi(x,t). \tag{17}$$

After inserting the plane wave function of the form: $\Psi(x,t)=e^{i(\pm k x-\frac{2\Delta Et}{h})}$,
into the time-dependent Schrödinger equation one get for $\Delta E$ :

$$\Delta E=\frac{h\hbar}{4\sqrt{M_1 M_2}}k^2. \qquad (18)$$

For $k = 2\pi/d = \pi/(a - \varepsilon)$ and $\Delta E=k_B T_c$ one get for the $T_c$:

$$T_c=\frac{\pi h^2}{2k_B}\frac{1}{\sqrt{M_1 M_2}(a-\varepsilon)^2}. \qquad (19)$$

This is the same equation as Eq. (15), for $M_1 = M_2$ one get the Eq. (8). At this point it is important to point out that pairs of (ψ(x,t), ψ(-x,t)) are eigenstates of the time-dependent Schrödinger equation.

### 1.3. Isotope effect in superconductivity

According to Eq. (8) and (10) , $T_c$ is proportional to 1/M. On the first view this is confusing, because one get a value of 1 for isotope effect exponent, which is twice larger then the expected value of 0.5. However, when isotopes are implanted into crystal, collisions between atoms with different masses, namely $M$ and $M_{iso}$ are present, and $T_{c(iso)}$ become proportional to $(MM_{iso})^{-1/2}$ (see Eq. 15). One can calculate the isotope effect exponent β from the following relation:

$$\beta=\frac{-\ln(\frac{T_c}{T_{ciso}})}{\ln(\frac{M}{M_{iso}})}=\frac{-\ln(\frac{\frac{A}{M}}{\frac{A}{\sqrt{MM_{iso}}}})}{\ln(\frac{M}{M_{iso}})}=\frac{-\ln(\frac{\sqrt{MM_{iso}}}{M})}{\ln(\frac{M}{M_{iso}})}=\frac{1}{2}\frac{\ln(\frac{M}{M_{iso}})}{\ln(\frac{M}{M_{iso}})}=0.5, \qquad (20)$$

where $A=\frac{\pi h^2}{2k_B}\frac{1}{(a-\varepsilon)^2}$. This calculation proves that the model of perfectly elastic collisions is also able to explain the isotope effect, which is one of the most relevant phenomena in superconductivity.

In binary, ternary, quaternary etc. compounds where two or more types of atoms build the crystal the isotope effect exponent β can take smaller values than 0.5. This is because p.e.c. between different types of atoms coexist, and thus also the dependency of the $T_c$ on mass varies greatly. Depending on succession of atom-atom collisions β can take more than one value, but experimentally it is possible to measure mostly only the smallest value, because the shift on the $T_c$ is minor for smaller β.

To elucidate this in more details, the β for binary compound of $MgB_2$ is calculated and compared with experimental values.

Similarly to the case 2H-$NbSe_2$ where Eq. (8) (*i.e.* collisions between atoms of the same sort are considered) and Eq. (15) (*i.e.* collisions between atoms of different sorts are considered) yield equal values for $T_c$ also in $MgB_2$. In the Table 1 the $T_c$ for $MgB_2$ was calculated by assuming that only Boron-Boron collisions contribute to the superconductivity and $\varepsilon = R_B/2$. Now if one consider that all collisions in the unit of atomic array of Mg-B-B are perfectly elastic the formula for $T_c$ transforms into:

$$T_c = \frac{\pi h^2}{2k_B} \frac{1}{\sqrt{M_B \sqrt{M_B M_{Mg}}}(a-\varepsilon)^2} = 39.5K \tag{21}$$

where $M_B$ and $M_{Mg}$ are atomic masses of Boron and Magnesium, respectively. While during Mg-B collisions the $\varepsilon_{Mg\text{-}B}$ is $\varepsilon_{Mg\text{-}B} = (R_{Mg}/2 + R_B/4) = 0.9625$ *Å,* during B-B collisions the $\varepsilon_{B\text{-}B}$ is $\varepsilon_{B\text{-}B} = (R_B/4 + R_B/4) = R_B/2$ (see Table 1). The mean value is $\varepsilon = (\varepsilon_{Mg\text{-}B} + \varepsilon_{B\text{-}B})/2 = 0.6937$ *Å.* After inserting these values for atomic masses, $\varepsilon$ and a=3.082 *Å* one get for $T_c$ a value of 39.5K. This indicates that not only Boron but also Magnesium participate in superconductivity. After implanting Boron isotopes into $MgB_2$ crystal, the atomic array Mg-B-B transforms into two possible arrays, namely Mg-B-$B_{iso}$ and Mg-$B_{iso}$-B, and their masses based on kinetic energy conservation are $\sqrt{M_{B(iso)}\sqrt{M_B M_{Mg}}}$ and $\sqrt{M_B \sqrt{M_{B(iso)} M_{Mg}}}$, respectively. By inserting these masses into the equation for isotope effect exponent β one get 0.5 and 0.25. The second value is in very good agreement with the experimental value of 0.26±0.03.

In the case of Magnesium isotope effect there are also two possible arrays, namely $Mg_{iso}$-B-B-Mg-B-B and Mg-B-B-$Mg_{iso}$-B-B. Based on the kinetic energy conservation their corresponding masses are $\sqrt{M_B\sqrt{M_B\sqrt{M_{Mg}\sqrt{M_B\sqrt{M_B M_{Mg(iso)}}}}}}$ and $\sqrt{M_B\sqrt{M_B\sqrt{M_{Mg(iso)}\sqrt{M_B\sqrt{M_B M_{Mg}}}}}}$, respectively. Inserting into $\sqrt{M_B\sqrt{M_B\sqrt{M_{Mg}\sqrt{M_B\sqrt{M_B M_{Mg(iso)}}}}}}$ the Eq. for β one get a value of 0.031 which is very close to the experimental value of 0.02.

### 1.4. The state of the pairing wave vectors (*k*, -*k*)

The quantum states of superconductivity and magnetic order coexist in many systems. Because of this, the Cooper pair formation is independent from the spin degrees of freedom. For instance in the heavy fermion system of $UGe_2$ where ferromagnetism (see Table 7) and superconductivity (see Table 1) coexist, is clear that spin-singlet Cooper pairs formation is impossible.

As it was pointed out above during perfectly elastic atom-atom collisions the kinetic energy is conserved forever. In other words the p.e.c. which are going away (k) must return back (-k), otherwise the kinetic energy is dissipating into heat. Now it is very simple to demonstrate the formation of Cooper pairs. Let suppose one start to measure the time at moment $t = t_0$ at any temperature $T < T_c$. The kinetic energy at the time $t_0$ is denoted as $E_k \propto k^2$, and the kinetic energy at any time $t > t_0$ is denoted as $E_{k'} \propto k'^2$. Which values can take the wave vector $k'$ at any time $t > t_0$? Based on the kinetic energy conservation one get: $k'^2 - k^2 = 0, \Rightarrow k' = \pm k$.

Following from the conservation of the kinetic energy, the connection between the wave vectors of the conduction charges ($k_{c.c.}$) and the phonon wave vectors ($k$) is given by the formula $k_{c.c.} = (m_e/M)^{1/2} k$ (see subsection 1.5.). Because of this linear dependence between $k_{c.c.}$ and $k$ the Cooper pair formation for conduction charges ($k_{c.c.}$, $-k_{c.c.}$) can be proved in the same way.

In this simple but tremendously important way it has been proved the symbiotic relationship between the state of the pairing wave vectors (*i.e.* Cooper pairs) ($k$, $-k$), ($k_{c.c.}$, $-k_{c.c.}$) and the p.e.c..

### 1.5.The incorporation of the conduction charges into the atom-atom p.e.c. through the London penetration depth at 0 K

In this subsection the incorporation of the conduction electrons or holes into the atom-atom p.e.c. will be treat. Below $T_c$ the electrical resistivity jumps at zero. When electrical resistivity is zero, then the kinetic energies of conduction electrons (holes) do not dissipate into heat. Now, one may ask, which are these conduction charges? The answer is that all conduction charges with a kinetic energy $E_{kc.c}$ that is equal to the kinetic energy transferred by the perfectly elastic atom-atom collisions contribute to the superconductivity.

This can be written as:

$$\frac{(hk_{c.c.})^2}{2m_e} = \frac{(hk)^2}{2M}. \qquad (22)$$

where $m_e$ and $M$ are the electron and atomic masses, respectively. From the Eq. (22) one get the linear dependence between $k_{c.c.}$ and $k$, which is $k_{c.c.} = (m_e/M)^{1/2} k$. For $k = 2\pi/2(\mathbf{a} - \varepsilon)$ and $k_{c.c.} = 2\pi/\lambda$ get the formula for the wave length, which is:

$$\lambda = 2(a - \varepsilon)\sqrt{\frac{M}{m_e}}. \qquad (23)$$

In the Table 4 is demonstrated that in many superconducting materials the $\lambda$ is of the same order as the London penetration depth at 0 K, *i.e.* $\lambda = \lambda_L(0)$.

| Material | M | $(a - \varepsilon)$(Å) | $\lambda_L(0)_{calc.}$(Å) | $\lambda_L(0)_{exp.}$(Å) |
|---|---|---|---|---|
| $MgB_2$ | $M_B$ | 2.657 | 754 | 850 [61]; 600 [62] |
| $OsB_2$ | $M_{Os}$ | 2.777 | 3282 | 3700 [63] |
| $MgCNi_3$ | $M_{Ni}$ | 2.47 | 1621 | 1280-1800 [64] |

Continuation of Table 4.

| | | | | |
|---|---|---|---|---|
| $LuNi_2B_2C$ | $M_{Ni}$ | 1.771 | 1162 | ≈1300 [65] |
| $PrOs_4Sb_{12}$ | $(M_{Os}M_{Sb})^{1/2}$ | 3.2775 | 3464 | 3440 [66] |
| $ZrB_{12}$ | $M_B$ | 6.975 | 1962 | ≈1550 [67] |
| CaAlSi | $M_{Si}$ | 3.639 | 1586 | $870_c$; $2060_{ab}$;[68] |
| 2H-$NbSe_2$ | $M_{Nb}$ | 1.993 | 1646 | 1250-1600 [69] |
| $Li_2Pd_3B$ | $M_B$ | 6.328 | 1782 | 1900 [32] |
| $Li_2Pt_3B$ | $(M_{Pt}M_B)^{1/2}$ | 4.98 | 2891 | 3640 [32] |
| $La_3Pd_4Ge_4$ | $M_{Ge}$ | 3.76 | 2746 | 2480 [43] |
| $CeCoIn_5$ | $M_{In}$ | 3.063 | 2812 | 2810 [70] |
| $La_3Pd_4Si_4$ | $M_{La}$ | 3.245 | 3277.4 | 3760 [44] |
| $CeCoIn_5$ | $M_{Ce}$ | 2.763 | 2803 | 2810 [70] |
| $UPd_2Al_3$ | $M_U$ | c-direction 2.435 | 3220 | 4500-4800 [71] |
| $UPd_2Al_3$ | $M_U$ | a-direction 3.6 | 4760 | 4500-4800 [71] |
| $UPt_3$ | $M_U$ | 4.697 | 6200 | ≈5200 [72] |
| URhGe | $M_U$ | 6.635 | 8771 | ≈9000 [73] |

*Table 4. Calculated $\lambda_L(0)$ for different superconducting materials are compared with experimental values. Except for $UPd_2Al_3$ in a and c-directions all values for $(a-\varepsilon)$ are taken from the Table 1.*

It is interesting the case of $UPd_2Al_3$ because in this compound seems that at lower temperatures in addition to the U-U p.e.c. in c-direction (see the last paragraph in the subsection 2.1.) also U-U collisions in a-direction become perfectly elastic and contribute to the superconductivity.

For superconducting materials such as Nb, Pb, Sn, Al and Cd the wave length of the conduction $\lambda$ is taken to be equal to $4*\lambda_L(0)$, i.e. $\lambda= 4*\lambda_L(0)$. For these materials the calculated values for $\lambda_L(0)$ are are given in the Table 5.

| Material | $(a-\varepsilon)(Å)$ | $\lambda_L(0)_{calc.}(Å)$ | $\lambda_L(0)_{exp.}(Å)$ |
|---|---|---|---|
| Pb | 1.35 | 416 | 390 |
| Sn | 2.457 | 573 | 510 |
| Al | 2.796 | 311 | 490 |
| Nb | 1.85 | 382 | 390 |
| Cd | 4.843 | 1100 | 1100 |

*Table 5. For $\lambda= 4*\lambda_L(0)$ the calculated values for the London penetration depth at 0 K are compared with the experimental results.*

In general results from different experimental methods yield different values for $\lambda_L(0)$. Considering this fact one can claim that the theoretical predictions are in good agreement with the experimental results.

It is very important to mention that from the linear dependence between $k_{c.c.}$ and $k$ one get also the linear dependence between $T_c$ and $\lambda_L(0)^{-2}$. This fact is in agreement with the experimental results from the Reference [89], where pressure effects on the penetration depth have been studied.

Based on the Eq. (23) one can specify the condition to be fulfilled by conduction charges that undergo the p.e.c. with the atoms . This condition is given as:

$$m_e \lambda^2 = 4M(a-\varepsilon)^2. \qquad (24)$$

For $\lambda = 2*\lambda_L(0)$ the Eq. (24) is transformed into:

$$m_e \lambda_L(0)^2 = M(a-\varepsilon)^2. \qquad (25)$$

To elucidate in more details the incorporation of the conduction charges into the atom-atom p.e.c. through the $\lambda_L(0)$ one schematic representation can be attached. This schematic representation is done for $\lambda = 2*\lambda_L(0)$ and $\lambda_L(0) = 4*a$, where a is the lattice constant. The value of $\lambda_L(0) = 4*a$ is taken only for a better graphical representation, and it has nothing to do with the experimental values, because it is two to three orders of magnitude smaller (see Tables 4 and 5).

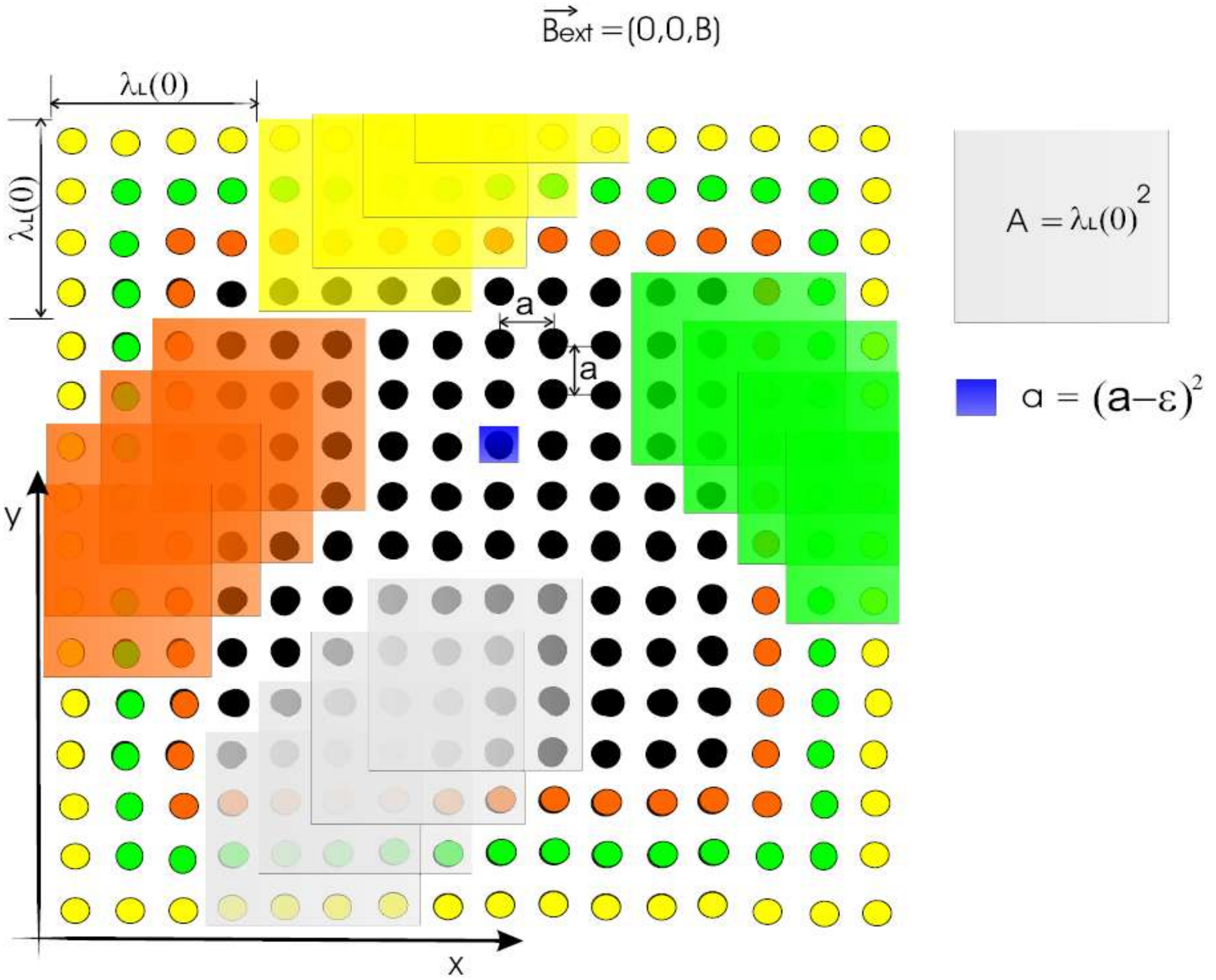

*Figure 2: An schematic representation for a 2D crystal with the lattice parameter a and $\lambda_L(0)=4a$. In the region of the black circles the Eq. (25) is fulfilled four times (i.e. for $B_{ext}<B_{c1}$ there are only superconducting electrons and the $B_{ext}$ can not penetrate into the crystal). In the red, green and yellow regions the Eq. (25) is fulfilled less than four times (red 3 times, green 2 times and yellow 1 time) and therefore in these regions the normal conducting and superconducting electrons coexist, i.e. the $B_{ext}$ penetrate into the crystal.*

In the Fig. 2 four different regions are depicted with yellow, green, red and gray colors. In the yellow region the condition (see Eq. 25) is fulfilled only once (*i.e.* these conduction charges are in the superconducting state), three other times the condition (25) is not fulfilled (*i.e.* these conduction charges are in the normal conducting state). Therefore, in the yellow region the incident magnetic field $B_{ext}$<$B_{c1}$ is partially reflected. In the green region the Eq. (25) is twice fulfilled (superconducting electrons) and twice unfulfilled (normal conducting electrons). Hence in the yellow region of the incident magnetic field is reflected more than in the yellow region. In the red region the Eq. (25) is three times fulfilled and once unfulfilled . In this region most of the electrons are in superconducting state, but there are still some normal conducting electrons, and a small part of the $B_{ext}$ is reflected. In the gray region( black circles ) the Eq (25) is fulfilled four times, and therefore all electrons are in superconducting state and this region the $B_{ext}$ *can not penetrate into the crystal.*

In the next paragraphs the upper critical magnetic field $B_{c2}(0)$ is calculated without taking into account the anisotropy of the $B_{c2}(0)$.

The equation for $B_{c2}$ is derived from the assumption that when the magnetic energy stored in the volume Ω that is of the order of $(\boldsymbol{a}-\varepsilon)^3$, is equal to the kinetic energy transported during atom-atom p.e.c. then the superconducting state is destroyed. One can write this as:

$$B_{c2}^2(0)\frac{\Omega}{2\mu_0}=\frac{\pi^2 h^2}{2M}\frac{1}{(a-\varepsilon)^2} \tag{26}$$

where $\mu_0$ is the magnetic permeability for vacuum.

As one can see the from the Table 6 there is an amazing agreement between calculated and experimental data. However, it is important to point out that for many component compounds it is not very simple to identify which atoms are active at low temperatures. Also because of the availability of the local magnetic fields and anisotropy of $B_{c2}(0)$, the determination of the volume Ω needs careful analysis. It is interesting that for the minimal and the maximal calculated $B_{c2}(0)$ values of 0.4 T ($ZrZn_2$) and 227 T ( $YBa_2Cu_3O_7$ ) the volume Ω is taken to be equal to $(\boldsymbol{a}-\varepsilon)^3/4$.

| Material | M | $(a-\varepsilon)$(Å) | Ω | $B_{c2}(0)$cal. (T) | $B_{c2}(0)$ exp.(T) |
|---|---|---|---|---|---|
| $MgB_2$ | B | 2.657 | $(\boldsymbol{a}-\varepsilon)^3$ | 15 | 14-18 [74] |
| $Sr_2RuO_4$ | Sr | 4.47 | $(\boldsymbol{a}-\varepsilon)^3$ | 1.44 | 1.5 [75] |
| $YNi_2B_2C$ | B | 3.101 | $(\boldsymbol{a}-\varepsilon)^3$ | 10.2 | 10 [76] |
| $YNi_2B_2C$ | C | 3.176 | $(\boldsymbol{a}-\varepsilon)^3$ | 9.15 | 10 [76] |
| $PrOs_4Sb_{12}$ | $(OsSb)^{1/2}$ | 3.2775 | $(\boldsymbol{a}-\varepsilon)^3$ | 2.37 | ≈2.37 [77] |
| 2H-$NbSe_2$ | Nb | 1.993 | $(\boldsymbol{a}-\varepsilon)^3$ | 10.5 | 5.3-17.3 [78] |

Continuation of Table 6.

| | | | | | |
|---|---|---|---|---|---|
| $UPd_2Al_3$ | U | 2.435 | $(a-\varepsilon)^3$ | 3.9 | 3.9 [79] |
| $UGe_2$ | U | 3.241 | $(a-\varepsilon)^3$ | 1.95 | 1.9 [80] |
| $YBa_2Cu_3O_7$ | $(CuO)^{1/2}$ | (1.93-0.975) | $(a-\varepsilon)^3/4$ | 227 | ≈250 [81] |
| $Li_2Pd_3B$ | B | 6.328 | $(a-\varepsilon)^3/4$ | 3.44 | 3.66 [82] |
| $Li_2Pt_3B$ | $(PtB)^{1/2}$ | 4.9802 | $(a-\varepsilon)^3/4$ | 3.04 | -- |
| $LiTi_2O_4$ | O | 3.6 | $(a-\varepsilon)^3/4$ | 11.6 | 11.6 [83] |
| URhGe | U | 6.635 | $(a-\varepsilon)^3/4$ | 0.65 | 0.71[80] |
| $ZrZn_2$ | Zr | 9.68 | $(a-\varepsilon)^3/4$ | 0.4 | 0.4 [[84]] |
| $KOs_2O_6$ | O | 2.219 | $(2^{1/2}/3)*(a-\varepsilon)^3$ | 28.3 | ≈33 [85] |
| $RbOs_2O_6$ | $(RbO)^{1/2}$ | 3.582 | $(2^{1/2}/3)*(a-\varepsilon)^3$ | 5.6 | ≈6 [86] |
| $Ba_{0.6}K_{0.4}BiO_3$ | $M_O$ | 2.355 | $1/2*[(2^{1/2}/3)*(a-\varepsilon)^3]$ | 34.49 | 32 at 1.8 K [87] |
| $MgCNi_3$ | $M_{Ni}$ | 2.47 | $1/2*[(2^{1/2}/3)*(a-\varepsilon)^3]$ | 15.9 | ≈14.4 [88] |

*Table 6 Calculations of $B_{c2}(0)$ for different superconducting materials are compared wit experimental data. For compounds with octahedral units the volume Ω is taken to be equal to* $(2^{1/2}/3)*(a-\varepsilon)^3$ *or the half of* $(2^{1/2}/3)*(a-\varepsilon)^3$.

## 2. Magnetic order

The collective quantum state where electronic magnetic moments are aligned parallel and anti-parallel to each other is called a state of magnetic order. Similar to superconductivity each magnetically ordered state is characterized by a critical temperature $T_c$ , which in the case of parallel alignment (↑↑) of magnetic moments (*i.e.*, ferromagnetic order) is called Curie-temperature and in the case of anti-parallel alignment (↑↓) of magnetic moments (*i.e.*, antiferromagnetic order) Nèel-temperature.

### 2.1. The method of the perfectly elastic collisions

In the following page an equation for calculating the $T_c$ for a magnetic ordered state will first be derived . Then, more for a didactic purpose, a schematic representation (see Fig. 3) illustrates how, due to perfectly elastic collisions, the state of magnetic order can be accomplished. And finally in Tables 7 and 8, the calculated values for $T_c$ are compared with experimental values.

During two p.e.c. of electrons with their neighboring atoms, each electron can travel a distance of $d = 2a$, where $a$ is the lattice parameter perpendicular to the orientation of the local magnetic moments. During four or six perfectly elastic collisions, electrons can travel distances of $d = 4a$ and $d = 6a$, respectively. To include all cases in one equation the distance $d$ is written as $d=2\kappa a$ where $\kappa$ takes values of 1, 2, 3, 4 *etc.*. If the electron velocity at the magnetically ordered state is $V$, then the time $\Delta t$ needed for an electron to travel distance $d$ is:

$$\Delta t=\frac{2\kappa a}{V}. \tag{27}$$

During perfectly elastic electron-atom collisions the kinetic energy is conserved; *i.e.*,

$$\frac{P_e^2}{2<m_{eff}>}=\frac{P_a^2}{2M_a} \tag{28}$$

where $<m_{eff}>$ and $M_a$ are the average electron effective mass and atomic mass, respectively. Similar to Cu-O collisions in H$T_c$ S, for $P_e=<m_{eff}>V$ and $P_a=hk$, or vice versa $P_e=hk$ and $P_a=M_a V$, Eq. (28) is transformed into:

$$\sqrt{<m_{eff}>M_a}\,V=hk, \tag{29}$$

where $k = 2\pi/d$. Inserting velocity $V$ from Eq. (29) into Eq. (27), for $\Delta t$ one get:

$$\Delta t=2\kappa^2\sqrt{<m_{eff}>M_a}\frac{a^2}{\pi h}. \tag{30}$$

In the time interval $\Delta t$ the electron spin do not flip between +h/2 and -h/2. From the time-energy uncertainty relation $\Delta t\, \Delta E = h/2$, where $\Delta E = k_B T_c$, one get the following equation for $T_c$ results:

$$T_c=\frac{\pi h^2}{4k_B}\frac{1}{\kappa^2\sqrt{<m_{eff}>M_a}\,a^2}. \tag{31}$$

As one can see in Eq. (31), the critical temperature depends on only the average effective mass of the electron, atomic mass and lattice parameter.

The saying goes that a picture tells a thousand words. Based on this dictum the author would like to familiarize the reader with the phenomenon of magnetic order in non-dissipative systems accomplished through perfectly elastic collisions. Figs. 3a and 3b depict ferromagnetic and antiferromagnetic order. To simplify the matter, the indistinguishability of particles is not taken into account, and particles are uncharged. The green and red particles have masses of $m_e$ and $<m_{eff}>$, respectively; the blue particles have a mass of $M_a$. It is also idealized that there are massless and non-dissipative filaments that connect the green particles with blue ones.

Let us observe what happens in a non-dissipative 1D system at temperatures below $T_c$: At time $t = t_0$ the red particles start to move with kinetic energy $P^2_e / 2<m_{eff}>$, and they hit the green ones, which gain the kinetic energy of $P^2_a / 2M_a$. Because these collisions are perfectly elastic according to Eq. (28), they are equal. After $2\pi$ -rotations the green particles throw the red ones to the next connected pair of green and blue particles, so the collision process between green and red particles will continue to infinity. At this point, it is important to reiterate that, during the time interval between two (ferromagnetic) and four (antiferromagnetic) collisions of red particles with green ones, the system of the ordered magnetic system is adiabatic; in other words, during the time $\Delta t$ there is no heat exchange between the system of ordered magnetic moments and environment.

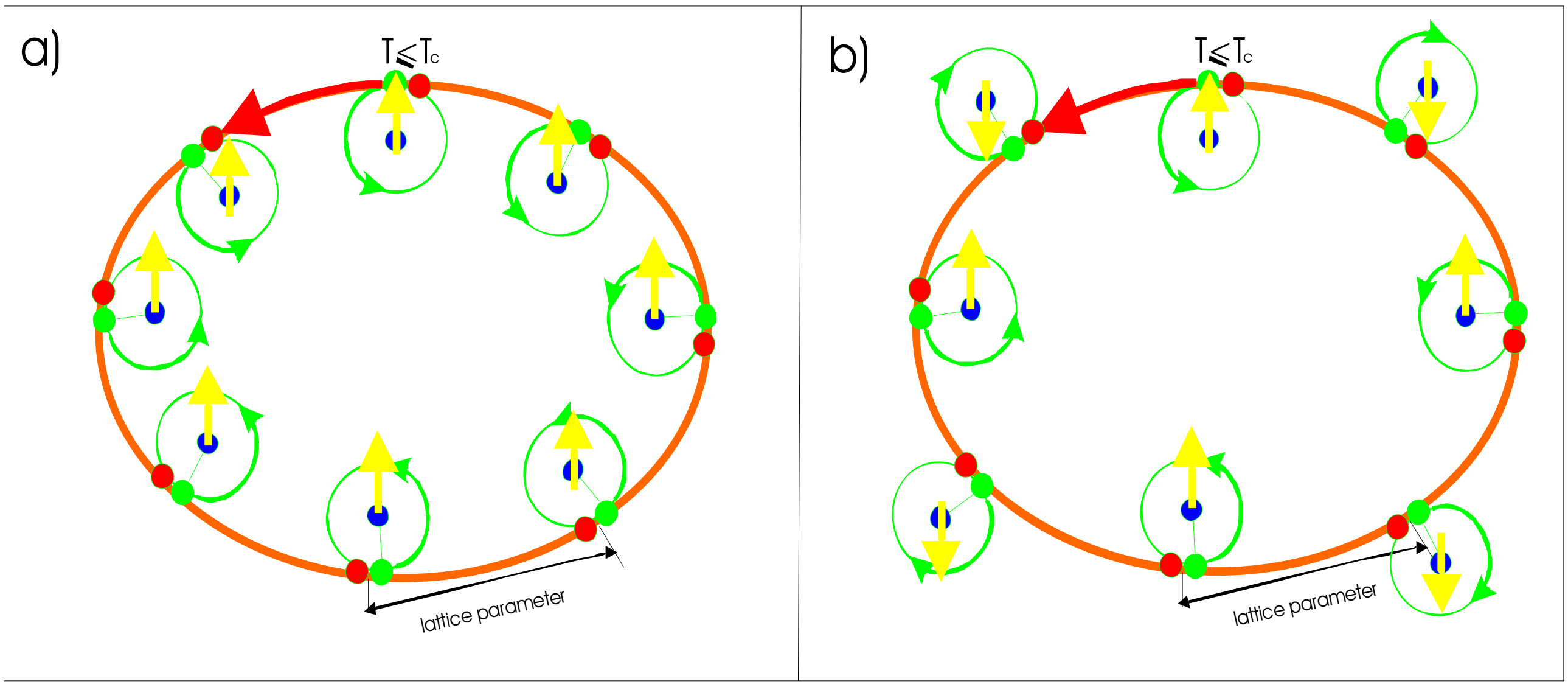

*Figure 3: Schematic representations for parallel (a) and anti-parallel (b) order of the non-dissipative 1D magnetic moments.*

Table 7 shows some calculated values for the critical temperatures for magnetically ordered states of different materials. The free-electron mass is denoted by $m_e$.

| Material, sruct. | $\kappa$ | $<m_{eff}>(m_e)$ | M | $a(\AA)$ | $T_c$ (K) calc. | $T_c$ (K) exp. |
|---|---|---|---|---|---|---|
| Fe, BCC ↑↑ | 1 | 1 | Fe | 2.866 | 1043 | 1043 [8] |
| Co, HEX ↑↑ | 1 | 1 | Co | 2.51 | 1323 | 1388 [8] |
| Co, HEX ↑↑ | 1 | 1 | Co | for min. Co-Co b. length of 2.45 [90] | 1388 | 1388 [8] |
| Ni, FCC ↑↑ | 1 | 1 | Ni | 3.52 | 674 | 627 [8] |
| Gd, HEX ↑↑ | 1 | 1.72 [128] | Gd | 3.64 | 293 | 293 [8] |
| $TiO_{(2-\delta)}$, ↑↑ | 1 | 1 | Ti | 0.7071*4.593 | 877 | 880 [91] |
| $GdFe_2Zn_{20}$, CUB ↑↑ | 1 | 1 | Fe | 0.7071*14.062 | 86 | 86 [92] |
| $GdAl_2$, $MgCu_2$ ↑↑ | 1 | 1 | Gd | 0.7071*7.899 | 163 | 170 [93] |

Continuation of Table 7.

| | | | | | | |
|---|---|---|---|---|---|---|
| $UGe_2$, ORTH. ↑↑ | 1 | ≈20 [94] at ambient-press. | U | $d_{U-U}$=(a+c)/2=4.05 [94] | 56 | 52 [94] |
| $UGe_2$, ORTH. ↑↑ | 1 | ≈60[94] at 15 (kbar) | U | $d_{U-U}$≈4.05 | 32 | 31 [95] |
| TbAl2, CUB. ↑↑ | 1 | 1 | $(Tb(TbAl)^{1/2})^{1/2}$ | a=7.8619 | 103 | 105 [96] |
| $EuFe_4$Sb12, ↑↑ | 1 | 1 | $(FeSb)^{1/2}$ | a=9.17 | 83.5 | 84 [97] |
| SrRuO3, ↑↑ | 1 | 1 | $(RuO)^{1/2}$ | $(2)^{1/2}$*b=7.82 | 164.6 | 165 [98] |
| ZrZn2, ↑↑ | 1 | 4.9 [50] | Zr | $(2)^{1/2}$*a=10.455 | 27.7 | 28 [50] |
| URhGe, ↑↑ | 1 | 47.6 [99] | U | c=7.51 | 10.6 | 9.5 [49] |
| $SmOs_4Sb_{12}$, ↑↑ | 1 | ≈170 | Sm | a=9.3085 | 2.3 | 2.6 [100] |
| $La_{0.45}Sr_{0.55}$MnO3,↑↑ ↑↓ A-type AFM | 1;2 | 1 | Mn; $(O(MnO)^{1/2})^{1/2}$ | $(a^2+c^2)^{1/2}$= 5.451; b =3.84 | $T_C$=291; $T_N$=232 | ≈300; ≈220 [101] |
| Cr, BCC ↑↓ | 2 | 0.73 [102] | Cr | 2.88 | 314 | 311 [8] |
| MnO, NaCl ↑↓ | 2 | 1 | Mn | 4.435 | 110 | 117 [103] |
| FeO, NaCl ↑↓ | 2 | 1 | Fe | 0.7071*4.334 | 228 | 198 [8] |
| $NpFeGa_5$, TET ↑↓ | 2 | 1 | Fe | a=4.2578 | 118 | 118 [104] |
| CoO, NaCl ↑↓ | 2 | 1 | Co | 0.7071*4.25 | 243 | 290 [8] |
| $UPt_3$, HCP ↑↓ | 2 | ≈170 | U | $d_{U-U}$=4.13 [105] | 4.5 | 5 [106] |
| $ScMnO_3$, ↑↓ | 2 | 1 | O | c/2=5.585 | 128.3 | 129 [107] |
| $YMnO_3$, ↑↓ | 2 | 1 | $(MnO)^{1/2}$ | a=6.139 | 78 | ≈80 [108] |
| $LaMnO_3$, ↑↓ | 2 | 1 | Mn | a=3.947 | 139 | 140 [109] |
| $LaTiO_3$, ↑↓ | 2 | 1 | Ti | c/2=3.958 | 147 | 146 [110] |
| $NdTiO_3$, ↑↓ | 2 | 1 | $(TiO)^{1/2}$ | b=5.659 | 94.5 | 94 [110] |
| $YTiO_3$, ↑↓ | 2 | 1 | Y | c=7.623 | 29 | 29 [110] |
| $LiFePO_4$, ORTH. ↑↓ | 2 | 1 | Fe | b=6.011 | 52 | 59 [111] |
| $NdFe_3(BO_3)_4$, HEX. ↑↓ | 2 | 1 | Fe | c=7.612 | 37 | 37 [112] |
| $Cu_2Fe_2Ge_4O_{13}$, ORTH. ↑↓ | 2 | 1 | $(FeO)^{1/2}$ | b=8.497 | 40.5 | ≈40 [113] |
| $Cu_2Fe_2Ge_4O_{13}$, ORTH. ↑↓ | 2 | 1 | $(CuO)^{1/2}$ | b=8.497 | 39.2 | ≈40[113] |
| $TbMnO_3$, ↑↓ | 2 | 1 | $(TbMn)^{1/2}$ | b=5.86 | 48 | 46 [114] |
| $TbMnO_3$, ↑↓ | 2 | 1 | Mn | c=7.36 | 39.8 | 41 [114] |

Continuation of Table 7.

| | | | | | | |
|---|---|---|---|---|---|---|
| Alpha-$Li_3Fe_2(PO_4)_3$ ↑↓ | 2 | 1 | Fe | a=8.562; c=8.616 | 29.3; 28.8 | 30 [115] |
| $Ni_2(TeO_3)_4Br2$, ↑↓ | 2 | 1 | Ni | c/2=8.154 | 31 | 29 [116] |
| $CuV_2O_6$, ↑↓ | 2 | 1 | Cu | a=9.168 | 24 | 24 [117] |
| Ca3Co2O6, HEX ↑↓ | 2 | 1 | Co | a=9.06 | 25.5 | 26 [118] |
| Na0.5CoO2, HEX. ↑↓ | 2 | 1 | Co | $(3)^{1/2}$*a=4.877 | 87.5 | 88 [119] |
| CuB2O4, ↑↓ | 2 | 1 | $(CoO)^{1/2}$ | a=11.528 | 21 | 21 [120] |
| $ZnCr_2Se_4$, ↑↓ | 2 | 1 | Cr | a=10.498 | 20 | 21 [121] |
| $LiCoPO_4$, ORTH. ↑↓ | 2 | 1 | Co | a=10.093 | 20.4 | ≈21 [122] |
| $LiNiPO_4$, ORTH. ↑↓ | 2 | 1 | Ni | a=10.0317 | 20.7 | 20.8 [123] |
| $LiVSi_2O_6$, MON. ↑↓ | 2 | 1 | V | a=9.634 | 24 | 24 [124] |
| $LiVGe_2O_6$, MON. ↑↓ | 2 | 1 | V | a=9.863 | 23 | 23 [125] |
| $BaCu_2Si_2O_7$, ORTH. ↑↓ | 2 | 1 | $(CuBa)^{1/2}$ | b=13.178 | 9.5 | 9.2 [126] |
| GdCo2Zn20, CUB. ↑↓ | 2 | 1 | Co | $(2)^{1/2}$*a=19.175 | 5.66 | 5.7 [92] |
| $Ni_3Al$, $Cu_3Au$↑↑ | 4 | 1 | Ni | 3.568 | 41 | 41 [127] |

*Table 7. Calculated $T_c$-s for different materials are compared with experimental data.*

The purpose of the $T_c$-calculations listed in the Table 7 is to show the reliability of the model of perfectly elastic collision. Because of the large number of the calculated data, an separate discussion for each compound is impossible. However, in the next paragraph some points are discussed briefly.

By assuming an average electron effective mass of $\langle m_{eff} \rangle$ =1 ($m_e$) for Fe, Co, Ni, $GdAl_2$, MnO, FeO, CoO, $LiFePO_4$, Alpha-$Li_3Fe_2(PO_4)_3$, $LiCoPO_4$, $LiNiPO_4$, and $Ni_3Al$, one get $T_c$ values that are very close to experimental values. In the cases of Iron and Cobalt (for the minimal Co-Co bond length of 2.45 *Å* ) the agreement between calculated and experimental values for $T_c$-s is exact. The value of $\langle m_{eff} \rangle$ =1.72 ($m_e$) for Gadolinium is achieved by averaging the experimental cyclotron masses of 1.21, 1.19, 2.2 and 2.28, (see Ref. [128]). For Chromium the calculated values for effective magnetic moments on the surface $\mu_{eff}$ =2.57$\mu_B$, second layer $\mu_{eff}$ =0.94$\mu_B$, and bulk magnetic moment $\mu_{eff}$ =0.6$\mu_B$ are averaged, where for g factor of 2 the effective mass is given by $m_{eff} = \mu_B/\mu_{eff}$. Also the pressure dependence of $T_c$ for the heavy electron system $UGe_2$ can be explained with Eq. (31). The agreement between calculated and experimental values is also very good for the antiferromagnetic heavy electron system $UPt_3$ with $<m_{eff}> \approx 170$ ($m_e$).

In order to better understand the dynamics of p.e.c. in magnetically ordered materials, it is of tremendous importance to calculate $T_c$-s for the double perovskites $A_2BB'O_6$ (A = Ca, Sr, Ba; B=Fe, Cr; and B' = Re, Mo, W). In most of these materials perfectly elastic atom-atom and atom-electron collisions coexist. Because of this coexistence Eq. (31) is transformed into:

$$T_c = \frac{\pi h^2}{4 k_B} \frac{1}{\kappa^2 \sqrt{(\sqrt{M_B \sqrt{M_A M_{B'}}}) < m_{eff} > a^2}}. \qquad (32)$$

In the cases of $Ca_2FeMoO_6$, and $Ca_2CrWO_6$ where $(M_{Ca} + M_{Fe}) = M_{Mo}$ and $2(M_{Ca} + M_{cr}) = M_W$, the cation groups of CaFeMo and CaCrW behave as if they are one cation with the mass of the heaviest cation multiplied by a factor of $(\sqrt{\frac{M_{B'}}{M_B}\sqrt{\frac{M_{B'}}{M_A}}})$ (this factor is derived by taking into account that all cation-cation collisions are perfectly elastic) and by the maximal velocity of the cation with the smallest mass. Considering this, Eq. (28) for kinetic energy conservation becomes:

$$\frac{P_e^2}{2 < m_{eff} >} = \frac{1}{2} M_{B'} (\sqrt{\frac{M_{B'}}{M_B}\sqrt{\frac{M_{B'}}{M_A}}}) V_A^2. \qquad (33)$$

$T_c$ values for these two materials are computed using the following equation:

$$T_c = \frac{\pi h^2}{4 k_B} \frac{1}{\kappa^2 (\sqrt{M_{B'} (\sqrt{\frac{M_{B'}}{M_B}\sqrt{\frac{M_{B'}}{M_A}}}) < m_{eff} > }) a^2}. \qquad (34)$$

The double perovskite of $Sr_2CrReO_6$ has the highest $T_c$ of 635 K [129]. In this compound the onset of the magnetic order can be explained only by assuming that there are no perfectly elastic cation-cation collisions. Because of this fact, Eq. (31) is used to calculate the $T_c$, where solely perfectly elastic atom-electron collisions are considered.

Calculated values for $T_c$ compared with the experimental data for the double perovskites $A_2BB'O_6$ are listed in Table 8. For all these materials the distance $a$ is taken to be equal to $\sqrt{8} R_A$ where: $R_{Ca}$=1,34 Å, $R_{Sr}$=1.44 Å and $R_{Ba}$=1.61 Å [130].

| Material | $\kappa$ | $<m_{eff}>(m_e)$ | *Eq.* | $T_c$ (K) calcul. | $T_c$ (K) exp. |
|---|---|---|---|---|---|
| $Ca_2FeReO_6$ ↑↑ | 1 | 1 | (32) | 535 | ≈538 |

Continuation of Table 8.

| | | | | | |
|---|---|---|---|---|---|
| $Sr_2FeReO_6$ ↑↑ | 1 | 1 | (32) | 419 | ≈403 |
| $Ba_2FeReO_6$ ↑↑ | 1 | 1 | (32) | 317 | ≈316 |
| $Ca_2FeMoO_6$ ↑↑ | 1 | 1 | (34) | 356 | 345-365 |
| $Sr_2FeMoO_6$ ↑↑ | 1 | 1 | (32) | 456 | 420-473 |
| $Ba_2FeMoO_6$ ↑↑ | 1 | 1 | (32) | 345 | ≈337 |
| $Ca_2CrWO_6$ ↑↑ | 1 | 1 | (34) | 166 | ≈161 |
| $Sr_2CrWO_6$ ↑↑ | 1 | 1 | (32) | 428 | ≈453 |
| $Sr_2CrReO_6$ ↑↑ | 1 | 0.73 [102] | (31) | 631 | ≈635 |

*Table 8. Calculated $T_c$-s for different double perovskites are compared with experimental values.*

It is evident that the agreement between calculated and experimental data is excellent.

In the context of coexistence of perfectly elastic atom-atom and atom-electron collisions, it is of particular interest to consider the ternary compound of $UPd_2Al_3$. $UPd_2Al_3$ is a heavy fermion system with electron effective mass of $<m_{eff}> \approx 50$ ($m_e$) [131], [132], and it shows coexistence of magnetic order ↑↓ and superconductivity, with $T_c$-s of 14,3 K and 2 K , respectively [133]. The hexagonal structure of this compound has lattice constants of $a$=5.350 Å and $c$=4.185 Å [134]. To get an exact agreement between experimental and calculated values for $T_c$ of magnetic order it is necessary to assume the coexistence of perfectly elastic U-Al collisions with electron-U collisions. In this case Eq. (31) is transformed into:

$$T_c = \frac{\pi h^2}{4 k_B} \frac{1}{\kappa^2 \sqrt{\sqrt{M_U M_{Al}} < m_{eff} >} a^2} . \qquad (35)$$

By inserting values: $<m_{eff}> = 50$ ($m_e$), $\kappa$ =2, $a = d_{U-U}$ = 4.185 Å and atomic masses for Uranium and Aluminum into Eq.(35) one get a value of 14.4 K for the Nèel temperature. To compute the critical temperature for superconductivity, Eq. (8) is used, where perfectly elastic U-U collisions are considered (*i.e.*, $M=M_U$), c= $d_{U-U}$ = 4.185 Å and $\varepsilon = R_U$ = 1.75 Å. After inserting these values into Eq. (8) one get for superconducting $T_c$ a value of 2.1K; that is in exact agreement with experimental value of 2K. In this case it is interesting to find out that only perfectly elastic U-U collisions contribute to the superconductivity, but not U-Al collisions which contribute to the magnetic order.

### 2.2The method based on the solution of the time-dependent Schrödinger equation

As it has been shown above the quantum state of magnetic order is caused due to the collisions between electrons and atoms with masses $m_e$ and $M$, and respective kinetic energies of $E_{ke}=p_e^2/2m_e$ and $E_{kM}=P_M^2/2M$. Because atom-electron collisions are perfectly elastic, their kinetic energies are conserved, i.e. $E_{ke}=E_{kM}$. In this case the corresponding kinetic term of the Hamiltonian operator for a free "particle" with the mass of $\sqrt{Mm_e}$ may be expressed as:

$$\hat{H}=\frac{-\hbar^2}{2\sqrt{Mm_e}}\frac{\partial^2}{\partial x^2}. \qquad (36)$$

The time-dependent Schrödinger wave equation is:

$$\frac{-\hbar^2}{2\sqrt{Mm_e}}\frac{\partial^2}{\partial x^2}\Psi(x,t)=i\hbar\frac{\partial}{\partial t}\Psi(x,t). \qquad (37)$$

After inserting the plane wave function of the form: $\Psi(x,t)=e^{i(\pm kx-\frac{4\Delta Et}{h})}$, into the time-dependent Schrödinger equation one get for $\Delta E$ :

$$\Delta E=\frac{h\hbar}{8\sqrt{Mm_e}}k^2. \qquad (38)$$

For $k=\pi/a$ and $\Delta E=k_B T_c$ one get for the $T_c$:

$$T_c=\frac{\pi h^2}{4k_B}\frac{1}{\sqrt{M m_e}a^2}. \qquad (39)$$

This is the same equation as Eq. (31), for $\kappa=1$. As one can see the pairs of (ψ(x,t), ψ(-x,t)) are eigenstates of the time-dependent Schrödinger equation.

## 2.3. Isotope effect in the magnetic order

According to the Eq. (1) and (2) the critical temperature is independent on atomic mass. However, the experimental evidence of the oxygen isotope effect in magnetically ordered system of $La_2CuO_4$ implies that critical temperature $T_c$ is dependent on the atomic mass. As one can see from the above equations, the model of the p.e.c. postulate a direct atomic mass dependence of $T_c$.

Before the Oxygen isotope effect in $La_2CuO_4$ is treated the $T_c$ -s for antiferromagnetic $La_2CuO_4$, $Ca_2RuO_4$, $Sr_2CuO_2Cl_2$, $Nd_2CuO_4$ and $Sr_2CuO_2Cl_2$ are calculated and compared with the experimental values. The crystal structures are orthorhombic, orthorhombic, tetragonal and tetragonal with lattice parameters in x-y plane: (a=5.358 *Å* , b=5.405 *Å* ), (a=5.4 *Å*, b=5.5 *Å* ), a=3.939 *Å* and a=3.974 *Å* respectively. If Oxygen atoms do not participate in the perfectly elastic collisions, *i.e.* only Cu-electron and Ru-electron collisions are considered, one get values of $T_c$*(in a-direction)*= 276.8 K, $T_c$*(in b-direction)*= 274.8 K for $La_2CuO_4$, $T_c$*(in diagonal-direction)*= 107 K for $Ca_2RuO_4$, $T_c$*(in diagonal-direction)*= 258 K for $Nd_2CuO_4$ and $T_c$*(in diagonal-direction)*= 254.2 K for $Sr_2CuO_2Cl_2$.These calculated values are very close with the experimental values for $La_2CuO_4$ which are between 256 K and 325 K, 112 K for $Ca_2RuO_4$, 245 K for $Nd_2CuO_4$ and 256.5 K for $Sr_2CuO_2Cl_2$. In all these calculations and in following calculations, the effective electron mass is taken to be equal to the free electron mass.

To get the hole spectra for $T_c$-s between 325 K and 257.4 K in $La_2CuO_4$ [135]-[136] it is necessary, in addition to the perfectly elastic Cu-electron collisions to include also the perfectly elastic Cu-O collisions. In the Table 9 are listed the calculated $T_c$-s for $La_2CuO_4$ in different directions in x-y plane:

| **direction (a,b)** | **(1,0)** | **(1,0)** | **(1,1/2)** | **(0.1)** | **(0.1)** | **(1/2,1)** |
|---|---|---|---|---|---|---|
| a ( *Å)* | 5.358 | 5.358 | 5.99 | 5.405 | 5.405 | 6.0319 |
| mass | M1 | $M_{Cu}$ | M2 | M1 | $M_{Cu}$ | M2 |
| $T_c(O^{16})$ | 328.8 | 276.8 | 315.9 | 326.5 | 274.8 | 311.5 |
| $T_c(O^{18})$ | 323.9 | 276.8 | 306.7 | 321.7 | 274.8 | 302.5 |
| $T_c(O^{16})$-$T_c(O^{18})$(K) | 4.8 | 0 | 9.1 | 4.78 | 0 | 9 |

*Table 9. Calculated $T_c$-s for* $La_2CuO_4$ *in different directions in x-y plane, where M1 and M2 are equal to* $\sqrt{M_{Cu}\sqrt{M_{Cu}M_O}}$ *and* $\sqrt{M_{Cu}M_O}$ *, respectively.*

Experimental results reveal values for $T_c(O^{16})$-$T_c(O^{18})$: 4.7 K, ~3 K, ~2 K and 0 K [135].

As one see from Table 9 the calculated values for $T_c(O^{16})$-$T_c(O^{18})$ are 9, 4.8 K and 0 K; the value of 2.6 K one can get for $T_c(O^{16})(0,1)$-$T_c(O^{18})(1,0)$). Except 9 K all three other values, namely, 4.8 K, 2.6 K and 0 K) for temperature shift are in very good agreement with the experimental results.

### 3. The coexistence between superconductivity and magnetic order

There are are many compounds in which the superconductivity and magnetic order coexist. As it has been shown above the model of p.e.c. is able to explain experimental results in many materials only if atom-atom and electron-atom collisions coexist. In the superconducting state the derivation of the formula for London penetration depth is based in electron-atom collisions. In many magnetically ordered systems (*e. g.* double perovskites, $UPd_2Al_3$ etc.) the coexistence between atom-atom and electron-atom collisions was necessary to be taken into account.

The coexistence between ferromagnetic and superconducting states is also possible in some heavy fermion systems such as: URhGe, $UGe_2$ and $ZnZr_2$. However, the peaceful coexistence between ↑↑ and superconducting states is feasible only if the strength of the internal magnetic field is smaller than the upper critical field $B_{c2}$ at temperatures below $T_c$, otherwise these two states can not coexist.

It is surprising that in the heavy fermion systems of URhGe (↑↑), $UPd_2Al_3$ (↑↓) and $UPt_3$ (↑↓), the electrons that participate into the magnetic order and superconductivity have different masses, namely the heavy electrons contribute to the magnetic order, and the light electrons with masses equal to to the free electron mass contribute to the superconductivity. As it has been shown above the $T_c$ calculations for magnetic order are in agreement with the experimental results only if the electron masses are taken to be of the order of $<m_{eff}> \approx 47.6$ ($m_e$), $<m_{eff}> \approx 50$ ($m_e$) and $<m_{eff}> \approx 170$ ($m_e$), for electrons in URhGe, $UPd_2Al_3$ and $UPt_3$, respectively. On the other hand in these compounds the London penetration depth calculations are in agreement with the experimental data (see Table 4) only if the effective electron masses are taken to be equal to the free electron mass, i.e. $<m_{eff}> = 1(m_e)$.

## 4. Conclusions

This paper presented new methods for calculating the critical temperatures of conventional superconductors, unconventional superconductors, ferromagnetic systems and antiferromagnetic sytems. Based on the model of p.e.c., the isotope effect in superconductivity and magnetic order has been also explained. The superb agreement between calculated values and experimental data for $T_c$ demonstrated the universality of the p.e.c. as the origin of collective quantum states.

For the quantum state of superconductivity it has been proved the symbiotic relationship between the state of the pairing wave vectors (*i.e.* Cooper pairs) ($k$, $-k$), ($k_{c.c.}$, $-k_{c.c.}$) and the p.e.c.. Also the incorporation of the conduction charges into the atom-atom p.e.c. through the London penetration depth at 0 K has been done. Based on the model of p.e.c. the upper critical field $B_{c2}$ at 0 K has been calculated for different superconducting materials.

In addition, the pressure dependence of $T_c$ for superconducting fcc Lanthanum, superconducting $YBa_2Cu_4O_8$ and for the ferromagnetic heavy electron system $UGe_2$ has been explained. From the results of the calculations it has been shown that in multiple element compounds such as, 2H-$NbSe_2$, $CeCoIn_5$, $YPd_2B_2C$, $ThPd_2B_2C$, $YNi_2B_2C$, $LuNi_2B_2C$ and $La_3Pd_4Si_4$ two or more types of atoms may contribute to the superconductivity.

In the Tables 7 and 8 a large number of $T_c$-calculations for different systems has been listed. For double perovskites $A_2BB'O_6$, the coexistence of perfectly elastic atom-atom with atom-electron collisions has been revealed. Also the calculations of the $T_c$-*s* for different 1D spin chains are done.

Finally, the coexistence of superconductivity and magnetic order (↑↑ and ↑↓ ) in different heavy fermion superconductors has been treated, where it has been shown that electrons that participate into the magnetic order and superconductivity have different masses, namely, the heavy electrons contribute to the magnetic order, and the light electrons contribute to the superconductivity.